\documentclass[12pt]{article}

\begin{document}

\hoffset = -0.3truecm
\voffset = -1.1truecm

\title{\bf
A Note on the Julia-Zee Dyon Solution\footnote{KEK Preprint: TH-1016; hep-th/0505058.}}

\author{
{\bf Rosy Teh\footnote{Permanent Address: School of Physics, Universiti Sains Malaysia, 11800 USM Penang, Malaysia}}\\
{\normalsize Institute of Particle and Nuclear Studies}\\
{\normalsize High Energy Accelerator Research Organization (KEK)}\\
{\normalsize Tsukuba, Ibaraki 305-0801, Japan}}

\date{May 2005}
\maketitle

\begin{abstract}
We observed that the Julia-Zee dyon solution can be presented in similar exact form when the $\phi$-winding number of the internal space is $n$.  However the closed form $n$-monopole version of the Julia-Zee dyon solution exits in the present of $(n-1)$ string antimonopoles. Hence the net monopole charge of the system at large distances is still unity. When $n=1$, the solution is just the Julia-Zee dyon solution. Using the ansatz of this solution, we also present the antimonopole version of the Julia-Zee dyon in closed form, with all the magnetic field directions reversed. We would also like to note that for a given monopole charge in this dyon solution, the net electric charge of the system can be both positive and negative.
\end{abstract}


\section{Introduction}
The SU(2) Yang-Mills-Higgs (YMH) field theory in $3+1$ dimensions, with the Higgs field in the adjoint representation had been shown to possess a large varieties of magnetic monopole configurations and hence dyon solutions. This is because it is always possible to construct a dyon solution from a static monopole solution.
The first well known monopole solution is the 't Hooft-Polyakov monopole solution \cite{kn:1} which is invariant under a U(1) subgroup of the local SU(2) gauge group. It has non zero Higgs mass and self-interaction. With the discovery of this numerical spherically symmetric monopole solution in 1974, Julia and Zee \cite{kn:2} in 1975 came up with a more general ansatz to construct the dyon solution from the given 't Hooft-Polyakov monopole solution. Both these solutions are numerical solutions. Immediately after the work of Julia and Zee \cite{kn:2}, Prasad and Sommerfield \cite{kn:3} gave the closed form solutions for both the 't Hooft-Polyakov monopole and Julia-Zee dyon. After Prasad and Sommerfield's work, Protogenov \cite{kn:4} presented the general form of the exact solution with two extra parameters. However this general solution was presented for the Yang-Mill sourceless equation.

There were various works that had attempted to generalize the Julia-Zee dyon solution \cite{kn:5}-\cite{kn:6}, one of which was the paper by I. Ju \cite{kn:5}. However this paper had several flaws. For example, the solution Eq.(15) given in their paper possessed a parameter $\delta$ which was not specified. Their solution Eq.(35c) is not a solution of the YMH equations. 

In general, configurations of the YMH field theory with a unit magnetic charge are spherically symmetric \cite{kn:1}-\cite{kn:6} although exceptions exist \cite{kn:7}. Multimonopole configurations with magnetic charges greater than unity possess at most axial symmetry \cite{kn:8}. It has been shown that these solutions cannot possess spherical symmetry \cite{kn:9}. 

So far exact monopole and multimonopoles solutions  \cite{kn:3}, \cite{kn:8} existed only in the Bogomol'nyi-Prasad-Sommerfield (BPS) limit. Outside this limit, when the Higgs field potential is non-vanishing, only numerical solutions are known. 
Recently, we have also shown that the ansatz of Ref.\cite{kn:10} possesses more exact multimonopole-antimonopole configurations in the BPS limit. We have also constructed the anti-configurations of all these multimonopole-antimonopole solutions \cite{kn:11} with all the magnetic monopole charges reversing their sign. Hence monopoles becomes antimonopoles and vice versa. 

Although the mathematics involved in the calculation of the dyon solution from a given monopole solution is standard in the SU(2) YMH field theory, the introduction of electric charge into the monopole systems is interesting and worth studying as it leads to more new physics. Also an exact BPS solution will become non-BPS with the introduction of an electric field in the monopole system. 

In this paper we would like to give some remarks on the exact Julia-Zee dyon solution \cite{kn:2}. We presented a $n$-monopole charge Julia-Zee dyon solution.  However this closed form $n$-monopole version of the Julia-Zee dyon solution exits only in the present of $(n-1)$ Dirac string antimonopoles which enter into the system in the form of a pure gauge potential linearly superposed with the non-Abelian gauge potential.  Hence the net monopole charge of the system is still unity at infinity. The relationship between the non-Abelian monopoles or (zeros of the Higgs field) and the Dirac string monopoles had been beautifully described by the work of Arafune et al. \cite{kn:12}. When $n=1$, the solution is just the Julia-Zee dyon solution. 

We also present the antimonopole version of the Julia-Zee dyon in closed form. The procedure involved been the same as those given in Ref.\cite{kn:11} for converting the multimonopole-antimonopoles configurations into the multi-antimonopole-monopoles configurations. Our study shows that with a given positive or negative monopole charge, the corresponding Julia-Zee dyon solutions can possess both positive or negative net electric charge, as to every given gauge potentials, the Higgs field can take on both $\pm$ signs. We would also like to comment on the general statement made by P. Rossi \cite{kn:13} (p.337) that the $\pm$ sign of the Bogomol'nyi equations, $(B^a_i \pm D_i \Phi^a) = 0$, corresponds to monopoles and antimonopoles respectively is not true. It is the $\phi$-winding number of the internal space that will have to be a positive or negative one in order to obtain a monopole or an antimonopole solution respectively.
Similarly the comment made by Marciano on p. 283 in Ref.\cite{kn:13a} and Marciano and Pagels on p. 244 in Ref.\cite{kn:13b} that the antimonopole can be obtained by changing the sign of the Higgs field is not correct. 

We briefly review the SU(2) Yang-Mills-Higgs field theory in the next section. In section 3, we present the general ansatz with $n$ $\phi$-winding number in the internal space coordinates system. We then carry on to present the $n$-monopole version of the dyon solution in the presence of $(n-1)$ Dirac string antimonopole and discussed some of its basic properties. In section 4, we obtained the $n$-antimonopole version of the solution and present the 1-antimonopole version of the dyon and its basic properties. We end with some comments and discussion in section 5.

\section{The SU(2) YMH Theory}
The SU(2) YMH Lagrangian in 3+1 dimensions with non vanishing Higgs potential is given by
\begin{equation}
{\cal L} = -\frac{1}{4}F^a_{\mu\nu} F^{a\mu\nu} + \frac{1}{2}D^\mu \Phi^a D_\mu \Phi^a - \frac{1}{4}\lambda(\Phi^a\Phi^a - \frac{\mu^2}{\lambda})^2. 
\label{eq.1}
\end{equation}

\noindent Here the Higgs field mass is $\mu$ and the strength of the Higgs potential is $\lambda$ which are constants. The vacuum expectation value of the Higgs field is $\mu/\sqrt{\lambda}$. The Lagrangian (\ref{eq.1}) is gauge invariant under the set of independent local SU(2) transformations at each space-time point.
The covariant derivative of the Higgs field and the gauge field strength tensor are given respectively by 
\begin{eqnarray}
D_{\mu}\Phi^{a} &=& \partial_{\mu} \Phi^{a} + g\epsilon^{abc} A^{b}_{\mu}\Phi^{c},\nonumber\\
F^a_{\mu\nu} &=& \partial_{\mu}A^a_\nu - \partial_{\nu}A^a_\mu + g\epsilon^{abc}A^b_{\mu}A^c_\nu.
\label{eq.2}
\end{eqnarray}
Since the gauge field coupling constant $g$ can be scaled away, we can set $g$ to one without any loss of generality. The metric used is $g_{\mu\nu} = (- + + +)$. The SU(2) internal group indices $a, b, c$ run from 1 to 3 and the spatial indices are $\mu, \nu, \alpha = 0, 1, 2$, and $3$ in Minkowski space.

The equations of motion that follow from the Lagrangian (\ref{eq.1}) are
\begin{eqnarray}
D^{\mu}F^a_{\mu\nu} &=& \partial^{\mu}F^a_{\mu\nu} + \epsilon^{abc}A^{b\mu}F^c_{\mu\nu} = \epsilon^{abc}\Phi^{b}D_{\nu}\Phi^c,\nonumber\\
D^{\mu}D_{\mu}\Phi^a &=& -\lambda\Phi^a(\Phi^{b}\Phi^{b} - \frac{\mu^2}{\lambda}),
\label{eq.3}
\end{eqnarray}
and the Bogomol'nyi equations is 
\begin{equation}
B^a_i \pm D_i \Phi^a = 0,
\label{eq.4}
\end{equation}
holds in the limit of vanishing $\mu$ and $\lambda$.
In the case of the exact BPS 't Hooft-Polyakov monopole solution, the $\pm$ sign does not correspond to monopoles and antimonopoles respectively as was stated in Ref.\cite{kn:13}. Instead the $\pm$ sign corresponds to a change in sign of the Higgs field. When the electric field is switched on, the changed in sign of the Higgs field will correspond to a change in sign of the electric field and hence the electric charge of the dyon solutions.
In the case of Ref.\cite{kn:10}, the multimonopole-antimonopoles  are solved with the $+$ sign, whereas the corresponding anti-multimonopole-monopole solutions \cite{kn:11} are solvable with the $-$ sign of the Bogomol'nyi equations together with a negative one $\phi$-winding number of the internal space. 

The tensor to be identified with the electromagnetic field, as was proposed by 't Hooft \cite{kn:1} is
\begin{eqnarray}
F_{\mu\nu} &=& \hat{\Phi}^a F^a_{\mu\nu} - \epsilon^{abc}\hat{\Phi}^{a}D_{\mu}\hat{\Phi}^{b}D_{\nu}\hat{\Phi}^c,\nonumber\\
	&=& \partial_{\mu}A_\nu - \partial_{\nu}A_\mu - \epsilon^{abc}\hat{\Phi}^{a}\partial_{\mu}\hat{\Phi}^{b}\partial_{\nu}\hat{\Phi}^c,
\label{eq.5}
\end{eqnarray}

\noindent where $A_\mu = \hat{\Phi}^{a}A^a_\mu$, the Higgs unit vector, $\hat{\Phi}^a = \Phi^a/|\Phi|$, and the Higgs field magnitude $|\Phi| = \sqrt{\Phi^{a}\Phi^{a}}$. 
The Abelian electric field is $E_i = F_{0i}$, and the Abelian magnetic field is $B_i = -\frac{1}{2}\epsilon_{ijk}F_{jk}$. 
We would also like to write the Abelian 't Hooft electromagnetic field as
\begin{equation}
F_{\mu\nu} = M_{\mu\nu} + H_{\mu\nu},
\label{eq.6}
\end{equation}

\noindent where 
\begin{eqnarray}
M_{\mu\nu} &=&  \partial_{\mu}A_\nu - \partial_{\nu}A_\mu, \nonumber\\
H_{\mu\nu} &=& - \epsilon^{abc}\hat{\Phi}^{a}\partial_{\mu}\hat{\Phi}^{b}\partial_{\nu}\hat{\Phi}^c,
\label{eq.7}
\end{eqnarray}

\noindent which we refer to as the gauge part and the Higgs part of the 't Hooft electomagnetic field respectively. 

The topological magnetic current \cite{kn:14} is defined to be 
\begin{equation}
k_\mu = \frac{1}{8\pi}~\epsilon_{\mu\nu\rho\sigma}~\epsilon_{abc}~\partial^{\nu}\hat{\Phi}^{a}~\partial^{\rho}\hat{\Phi}^{b}~\partial^{\sigma}\hat{\Phi}^c,
\label{eq.8}
\end{equation}

\noindent which is also the topological current density of the system and the corresponding conserved topological magnetic charge is
\begin{eqnarray}
M & = & \int d^{3}x~k_0 = \frac{1}{8\pi}\int \epsilon_{ijk}\epsilon^{abc}\partial_{i}\left(\hat{\Phi}^{a}\partial_{j}\hat{\Phi}^{b}\partial_{k}\hat{\Phi}^{c}\right)d^{3}x \nonumber\\
& = & \frac{1}{8\pi}\oint d^{2}\sigma_{i}\left(\epsilon_{ijk}\epsilon^{abc}\hat{\Phi}^{a}\partial_{j}\hat{\Phi}^{b}\partial_{k}\hat{\Phi}^{c}\right)\nonumber\\
& = & \frac{1}{4\pi} \oint d^{2}\sigma_{i}~B_i. 
\label{eq.9}
\end{eqnarray}

\noindent As mentioned by Arafune et al. \cite{kn:12}, the magnetic charge M is the total magnetic charge of the system if and only if the gauge field is non singular. If the gauge field is singular and carries Dirac string monopoles, then the total magnetic charge of the system is the total sum of the Dirac string monopoles and the monopoles carry by the Higgs field which is $M$.


\section{Solution with Positive $\phi$-Winding Number}
The magnetic ansatz of Ref.\cite{kn:10} can be further generalized in the standard way to accommodate for the presence of an electric field by writing 
\begin{eqnarray}
A^a_0 &=& \sinh\gamma\left(\frac{1}{r}\chi_1(r)\hat{u}^a_r + \frac{1}{r}\chi_2(\theta)\hat{u}^a_\theta\right)\nonumber\\
A_i^a &=&  - \frac{1}{r}\psi_1(r) \hat{u}^{a}_\phi\hat{\theta}_i + \frac{n}{r}\psi_2(r)\hat{u}^{a}_\theta\hat{\phi}_i
+ \frac{1}{r}R_1(\theta)\hat{u}^{a}_\phi\hat{r}_i - \frac{n}{r}R_2(\theta)\hat{u}^{a}_r\hat{\phi}_i \nonumber\\
\Phi^a &=& \cosh\gamma\left(\frac{1}{r}\chi_1(r)~\hat{u}^a_r + \frac{1}{r}\chi_2(\theta)\hat{u}^a_\theta\right),
\label{eq.10}
\end{eqnarray}
where $\gamma$ is an arbitrary constant.
\noindent The spherical coordinate orthonormal unit vectors, $\hat{r}_i, \hat{\theta}_i$, and $\hat{\phi}_i$ are defined by 
\begin{eqnarray}
\hat{r}_i &=& \sin\theta ~\cos \phi ~\delta_{i1} + \sin\theta ~\sin \phi ~\delta_{i2} + \cos\theta~\delta_{i3},\nonumber\\
\hat{\theta}_i &=& \cos\theta ~\cos \phi ~\delta_{i1} + \cos\theta ~\sin \phi ~\delta_{i2} - \sin\theta ~\delta_{i3},\nonumber\\
\hat{\phi}_i &=& -\sin \phi ~\delta_{i1} + \cos \phi ~\delta_{i2}.
\label{eq.11}
\end{eqnarray}

\noindent and the isospin coordinate orthonormal unit vectors, $\hat{u}_r^a, \hat{u}_\theta^a$, and $\hat{u}_\phi^a$ are defined by 
\begin{eqnarray}
\hat{u}_r^a &=& \sin\theta ~\cos n\phi ~\delta_{1}^a + \sin\theta ~\sin n\phi ~\delta_{2}^a + \cos\theta~\delta_{3}^a,\nonumber\\
\hat{u}_\theta^a &=& \cos\theta ~\cos n\phi ~\delta_{1}^a + \cos\theta ~\sin n\phi ~\delta_{2}^a - \sin\theta ~\delta_{3}^a,\nonumber\\
\hat{u}_\phi^a &=& -\sin n\phi ~\delta_{1}^a + \cos n\phi ~\delta_{2}^a.
\label{eq.12}
\end{eqnarray}

\noindent To solve for solutions, the ansatz (\ref{eq.10}) is substituted into the equations of motion (\ref{eq.3}) in the limit of vanishing $\lambda$, $\mu$ and $\mu^2/\lambda \rightarrow \beta^2\cosh^2\gamma$. We find that when we set 
\begin{eqnarray}
\psi_1(r)=\psi(r),~~~\chi_1(r)=\chi(r),~~~R_1(\theta)=\chi_2(\theta)=0,\nonumber\\
n\psi_2(r)=(n-1)+\psi(r),~~~nR_2(r)=(n-1)\cot\theta,
\label{eq.13}
\end{eqnarray}
the equations of motion can be simplified to the two usual second order differential equations,
\begin{eqnarray}
r^2\psi^{\prime\prime} - (1-\psi)(2\psi-\psi^2-\chi^2) &=& 0,\nonumber\\
r^2\chi^{\prime\prime} - 2\chi(1-\psi)^2 &=& 0,
\label{eq.14}
\end{eqnarray}
where prime means $\frac{d}{dr}$.

In this note, we would like to point out that the second order coupled non linear equations (\ref{eq.14}) can be integrated once to give two sets of first order coupled non linear differential equations,
\begin{eqnarray}
r\psi^{\prime} \pm \chi(1-\psi) &=& 0\nonumber\\
 r\chi^{\prime} - \chi \pm 2\psi \mp \psi^2 &=& 0. 
\label{eq.15}
\end{eqnarray}
These two sets of first order coupled differential equations (\ref{eq.15}) actually correspond to solving for the Bogomol'nyi equations, $(B^a_i \pm D_i\Phi^a) = 0$, when $\gamma = 0$ in ansatz (\ref{eq.10}).

The general solutions to this system of equations (\ref{eq.15}) that can be reduced to the Julia-Zee dyon solution are \cite{kn:4}
\begin{eqnarray}  
\psi(r) = \psi_{\pm} &=& 1 \pm \frac{\beta r}{(\sqrt{a^2 + 1}\sinh(\beta r) + a\cosh(\beta r))}\nonumber\\
\chi(r) = \chi_{\pm} &=& \pm \left(1 - \beta r \frac{(\sqrt{a^2 + 1}\cosh(\beta r) + a\sinh(\beta r))}{(\sqrt{a^2 + 1}\sinh(\beta r) + a\cosh(\beta r))}\right). 
\label{eq.16}
\end{eqnarray}
Here $a$ and $\beta$ are arbitrary constants. 
There are all together four different sets of solutions in Eq.(\ref{eq.16}), as to each $\psi_{\pm}$, there are two $\chi = \chi_{\pm}$ which arise from solving the two different sets of first order differential equations, Eq.(\ref{eq.15}). When $a$ is set to zero, solutions (\ref{eq.16}) become the exact Julia-Zee dyon solutions \cite{kn:3} and when $a = i$, they become the complex infinite energy dyon solutions discussed by Singleton \cite{kn:15}. The parameter $\beta$ determines the expectation value of the Higgs field as, $\Phi^a \rightarrow \mp(\beta\cosh\gamma)\hat{u}^a_r$, at $r$ infinity.

With Eq.(\ref{eq.13}), the gauge potentials (\ref{eq.10}) reduce to,
\begin{eqnarray}
A^a_0 &=& \sinh\gamma\frac{1}{r}\chi(r)~\hat{u}^a_r \nonumber\\
A_i^a &=&  \frac{1}{r}\psi(r)(\hat{u}^{a}_\theta\hat{\phi}_i - \hat{u}^{a}_\phi\hat{\theta}_i) + \frac{(n-1)}{r}\{\hat{u}^{a}_\theta-\cot\theta~\hat{u}^{a}_r\}\hat{\phi}_i\nonumber\\
\Phi^a &=& \cosh\gamma\frac{1}{r}\chi(r)~\hat{u}^a_r,
\label{eq.17}
\end{eqnarray}
which becomes the Julia-Zee dyon ansatz when $n=1$. The $\phi$-winding number, $n$, in the gauge potentials (\ref{eq.17}) is restricted to only positive integer. Hence the solutions obtained from this ansatz possess positive magnetic charge. 
We also notice that the pure gauge potential term of $A_i^a$ in Eq.(\ref{eq.17}) can be simplified to
\begin{equation}
A^a_i(\mbox{pure})=\frac{(n-1)}{r}\{\hat{u}^{a}_\theta-\cot\theta~\hat{u}^{a}_r\}\hat{\phi}_i = -(n-1)\delta^a_3 \partial_i\phi, 
\label{eq.18}
\end{equation}
and it possesses a line singularity. As this singular gauge potential is a pure gauge, it does not appear in the non-Abelian electromagnetic field $F^a_{\mu\nu}$. Hence calculating for the non-Abelian magnetic and the electric field from the gauge potentials (\ref{eq.17}), we get
\begin{eqnarray}
B^a_i &=&  \frac{1}{r^2}\{r\psi^{\prime}(\hat{u}^{a}_\theta\hat{\theta}_i + \hat{u}^{a}_\phi\hat{\phi}_i) + \psi(2-\psi)\hat{u}^{a}_r\hat{r}_i\}\nonumber\\
E_i^a &=&  -\frac{1}{r^2}\sinh\gamma\{(r\chi^{\prime}-\chi)\hat{u}^{a}_r\hat{r}_i + \chi(1-\psi)(\hat{u}^{a}_\theta\hat{\theta}_i + \hat{u}^{a}_\phi\hat{\phi}_i) \},
\label{eq.19}
\end{eqnarray}
which is just the exact Julia-Zee dyon magnetic and electric field in the $n$ $\phi$-winding number internal space. 
The energy or mass of the system is given by
\begin{eqnarray}
M &=& \int~dx^3 (\frac{1}{4}F^a_{ij}F^a_{ij} + \frac{1}{2}F^a_{0i}F^a_{0i} + \frac{1}{2}D_i\Phi^a D_i\Phi^a + \frac{1}{2}D_0\Phi^a D_0\Phi^a)\nonumber\\
 &=& \int~dx^3 (\frac{1}{2}B^a_{i}B^a_{i} + \frac{1}{2}E^a_{i}E^a_{i} + \frac{1}{2}D_i\Phi^a D_i\Phi^a)\nonumber\\
&=& \left.\frac{4\pi}{r}\cosh^2\gamma~\chi(r\chi^{\prime} - \chi) \right|^{r=\infty}_{r=0}.
\label{eq.20}
\end{eqnarray}
Therefore for solutions (\ref{eq.16}), the energy is finite, $M=4\pi\beta\cosh^2\gamma$, only when $a=0$.

From the ansatz (\ref{eq.10}), the Abelian 't Hooft gauge potentials are given by 
\begin{equation}
A_0 = \sinh\gamma \frac{1}{r}\chi(r), ~~~A_{i} = \hat{\Phi}^{a}A^{a}_{i} = -(n-1)\cos\theta~\partial_i \phi. 
\label{eq.21}
\end{equation}
Therefore the Abelian 't Hooft magnetic and electric fields are respectively,
\begin{eqnarray}
B_i &=&  M_i + H_i = -\frac{(n-1)}{r^2}\hat{r}_i + \frac{n}{r^2}\hat{r}_i = \frac{1}{r^2}\hat{r}_i\nonumber\\
E_i &=&  -\frac{1}{r^2}\sinh\gamma(r\chi^\prime - \chi)\hat{r}_i 
\label{eq.22}
\end{eqnarray}
The 't Hooft magnetic field in Eq.(\ref{eq.22}) always corresponds to that of a unit monopole field irrespective of the values of $n$ and $a$ in the solutions. When $n \not= 1$, the configuration (\ref{eq.17}) can be viewed as that of a $n$-monopole in the Higgs field superposed with a $(n-1)$ string antimonopole in the pure gauge field \cite{kn:12}. Therefore the net magnetic charge of the system is one at large $r$. Off course when $n=1$, the string antimonopole disappear. When $a=0$, the monopole in the Higgs field is the finite energy 't Hooft monopole. However when $a \not= 0$, the solution is singular at $r=0$ and the monopole in the Higgs field becomes a Wu-Yang type monopole with infinite energy. 

Similary, the 't Hooft electric field is non singular at the origin, $r=0$, when $a=0$. However when $a$ takes value other than zero, a point singularity exist at $r=0$. This can be seen by calculating for the total electric charge,
\begin{equation}
Q = \sinh\gamma \{-4\pi(r\chi^\prime - \chi)\delta(r) \pm \left.4\pi\psi(2-\psi)\right|^{r=\infty}_{r=0} \} = \pm 4\pi\sinh\gamma, 
\label{eq.23}
\end{equation}
for all values of $a$. The net electric charge comes from the delta function point source when $a\not= 0$ and when $a=0$, there is no delta function point source and the net charge is from the regular electric charge distribution, that is the second term of Eq.(\ref{eq.23}). We also notice that the net electric charge can be both positive or negative in a positive monopole charge field.


\section{Solution with Negative $\phi$-Winding Number}
Using the same procedure as in Ref.\cite{kn:11}, we can construct the antimonopole version of the Julia-Zee dyon by setting 
\begin{eqnarray}
\psi_1(r)=\psi(r),~~~\chi_1(r)=\chi(r),~~~R_1(\theta)=\chi_2(\theta)=0,\nonumber\\
n\psi_2(r)=(n+1)-\psi(r),~~~nR_2(r)=(n+1)\cot\theta,
\label{eq.24}
\end{eqnarray}
in the ansatz (\ref{eq.10}).
The gauge potentials of the anti-configuration then takes the form
\begin{eqnarray}
A^a_0 &=& \sinh\gamma\frac{1}{r}\chi(r)~\hat{u}^a_r \nonumber\\
A_i^a &=& - \frac{1}{r}\psi(r)(\hat{u}^{a}_\theta\hat{\phi}_i + \hat{u}^{a}_\phi\hat{\theta}_i) + \frac{(n+1)}{r}\{\hat{u}^{a}_\theta-\cot\theta~\hat{u}^{a}_r\}\hat{\phi}_i\nonumber\\
\Phi^a &=& \cosh\gamma\frac{1}{r}\chi(r)~\hat{u}^a_r,
\label{eq.25}
\end{eqnarray}
and the pure gauge term becomes
\begin{equation}
A^a_i(\mbox{pure})=\frac{(n+1)}{r}\{\hat{u}^{a}_\theta-\cot\theta\hat{u}^{a}_r\}\hat{\phi}_i = -(n+1)\delta^a_3 \partial_i\phi. 
\label{eq.26}
\end{equation}
Substituting the solution Eq.(\ref{eq.24}) into the equations of motion (\ref{eq.3}) in the BPS limit will give the same two coupled second order differential equations for $\psi$ and $\chi$ as in Eq.(\ref{eq.14}). Hence the solutions (\ref{eq.16}) apply for the gauge potentials and scalar Higgs field of Eq.(\ref{eq.25}). In this case, the $\phi$-winding number, $n$, in solution (\ref{eq.24}) is restricted to only negative integers and the net monopole charge of the configuration is negative one irrespective of the value of $n$. When $n=-1$, the pure gauge term (\ref{eq.26}) disappear and the solution is the regular Julia-Zee dyon solution with a negative one 't Hooft monopole charge. We would prefer to call this solution the Julia-Zee (JZ) anti-dyon solution.

Calculating for the non-Abelian magnetic and the electric field from the gauge potentials (\ref{eq.25}), we get
\begin{eqnarray}
B^a_i &=&  \frac{1}{r^2}\{r\psi^{\prime}(-\hat{u}^{a}_\theta\hat{\theta}_i + \hat{u}^{a}_\phi\hat{\phi}_i) - \psi(2-\psi)\hat{u}^{a}_r\hat{r}_i\}\nonumber\\
E_i^a &=&  -\frac{1}{r^2}\sinh\gamma\{(r\chi^{\prime}-\chi)\hat{u}^{a}_r\hat{r}_i + \chi(1-\psi)(\hat{u}^{a}_\theta\hat{\theta}_i - \hat{u}^{a}_\phi\hat{\phi}_i) \},
\label{eq.27}
\end{eqnarray}
which is the exact JZ anti-dyon magnetic and electric field in the negative $n$ $\phi$-winding number internal space. The electromagnetic fields (\ref{eq.27}) are regular when $a=0$. When $a>0$, a point singularity exits at $r=0$ in the solution and the 't Hooft-Polyakov antimonopole becomes the Wu-Yang type antimonopole. When $a<0$, in addition to the point singularity at $r=0$, a spherical shell singuarity is also present at $\tanh(\beta r) = -\frac{a}{\sqrt{a^2+1}}$.

Similarly, the energy or mass of the anti-dyon system is given by
\[M = \left.\frac{4\pi}{r}\cosh^2\gamma~\chi(r\chi^{\prime} - \chi) \right|^{r=\infty}_{r=0},\]
and it is finite, $M=4\pi\beta\cosh^2\gamma$ for all negative values of $n$, but only when $a=0$.

From the ansatz (\ref{eq.10}), the Abelian 't Hooft gauge potentials is given by 
\begin{equation}
A_0 = \sinh\gamma \frac{1}{r}\chi(r), ~~~A_{i} = \hat{\Phi}^{a}A^{a}_{i} = -(n+1)\cos\theta~\partial_i \phi,
\label{eq.28}
\end{equation}
and the Abelian 't Hooft magnetic and electric fields are respectively,
\begin{eqnarray}
B_i &=&  M_i + H_i = -\frac{(n+1)}{r^2}\hat{r}_i + \frac{n}{r^2}\hat{r}_i = \frac{-1}{r^2}\hat{r}_i\nonumber\\
E_i &=&  -\frac{1}{r^2}\sinh\gamma(r\chi^\prime - \chi)\hat{r}_i 
\label{eq.29}
\end{eqnarray}
The 't Hooft magnetic field in Eq.(\ref{eq.29}) always corresponds to that of a unit antimonopole field irrespective of the values of $n$ and $a$ in the solutions. When $n=-1$, the string monopole disappear and the solution is regular over all space. When $n \not= -1$, the configuration (\ref{eq.25}) can be viewed as that of a $|n|$-antimonopole in the Higgs field superposed with a $|n+1|$ string monopole in the pure gauge field \cite{kn:12}. Therefore the net magnetic charge of the system is negative one at large $r$. 
When $a=0$, the monopole in the Higgs field is the finite energy 't Hooft-Polyakov antimonopole. However when $a \not= 0$, the pole in the Higgs field becomes a Wu-Yang type antimonopole with infinite energy. 

Similary, the 't Hooft electric field is non singular at the origin, $r=0$, when $a=0$. However when $a$ takes value other than zero, a point singularity exist at $r=0$. When $a>0$, the total electric charge is given by,
\begin{equation}
Q = \sinh\gamma \{-4\pi(r\chi^\prime - \chi)\delta(r) \pm \left.4\pi\psi(2-\psi)\right|^{r=\infty}_{r=0} \} = \pm 4\pi\sinh\gamma, 
\label{eq.30}
\end{equation}
The net electric charge comes from the delta function point source when $a\not= 0$ and when $a=0$, there is no delta function point source and the net charge is from the regular electric charge distribution, that is the second term of Eq.(\ref{eq.30}). We also notice that the net electric charge can be both positive or negative in the presence of a negative monopole charge field. 


\section{Comments}
When solving the equations of motion (\ref{eq.3}), for solutions in the BPS limit, the $\pm$ sign of the Bogolmol' equations (\ref{eq.4}) is not significant. The second order coupled differential equations will always reduced to two different sets of first order coupled differential equations upon integration corresponding to the two sets of equations, $B^a_i \pm D_i\Phi^a$. The $\pm$ sign will only give a change in the sign of the corresponding Higgs field. When the electric field is switched on to give a dyon solution and $\gamma \not= 0$, the change in sign of the Higgs field will give a change in sign of the electric field and hence the charge present in the fields. The $\pm$ sign in the Bogomol'nyi equations will not change the sign of the monopole charge present in the fields. It is only the sign of the $\phi$-winding number, $n$, that will determines the presence of monopoles or antimonopoles.

Upon applying the usual gauge transformation,
\begin{eqnarray}
A^{\prime a}_{\mu} &=& A^{a}_{\mu}(\mbox{pure}) + \cos\theta A^a_\mu + \sin\theta \epsilon^{abc} A^b_\mu \hat{n}^c + 2\sin^2\frac{\theta}{2} \hat{n}^a (\hat{n}^bA^b_\mu),\nonumber\\
A^{a}_{\mu}(\mbox{pure})&=& \hat{n}^a\partial_\mu \theta + \sin\theta\partial_\mu\hat{n}^a + 2\sin^2\frac{\theta}{2}\epsilon^{abc} (\partial_\mu\hat{n}^b)\hat{n}^c
\label{eq.31}
\end{eqnarray}
on the gauge and Higgs field of Eq.(\ref{eq.17}), where $\hat{n}^a = \hat{u}^a_\phi$ to rotate the Higgs field from the $\hat{u}^a_r$ direction to the $\delta^a_3$ direction, we get,
\begin{eqnarray}
A^{\prime a}_{\mu} &=& \frac{1}{r}(1-\psi)(\hat{u}^a_\phi \hat{\theta}_i - \hat{u}^a_\rho \hat{\phi}_i)+ \frac{1}{r}\tan\frac{\theta}{2}\delta^a_3\hat{\phi}_i - (n-1)\delta^a_3\partial_i\phi,\nonumber\\
\Phi^{\prime a} &=& \frac{1}{r}\psi~\delta^a_3.
\label{eq.32}
\end{eqnarray}
From Eq.(\ref{eq.32}), we notice that the 't Hooft-Polyakov $n$-monopole in the Higgs field has been ``transfered" to the gauge field and a Dirac string one-monopole now exist in the gauge fields with no monopole in the Abelian Higgs field.

Therefore for the solutions (\ref{eq.16}) and (\ref{eq.17}), a 't Hooft-Polyakov $n$-monopole exist in the Higgs field and a string $(n-1)$-antimonopole is present in the pure gauge field (\ref{eq.18}), when $n>0$ and $a=0$. These solutions except for the presence of the pure gauge term are regular solutions and possess finite energy, electric and magnetic charge. The solutions can possess both positive and negative net electric charge. 

As for the solutions (\ref{eq.16}) and (\ref{eq.25}), a 't Hooft-Polyakov $|n|$-antimonopole exist in the Higgs field and a string $|n+1|$-monopole is present in the pure gauge field (\ref{eq.26}), when $n<0$ and $a=0$. Similarly, these solutions except for the presence of the pure gauge term are regular solutions and possess finite energy, electric and magnetic charge. Also in this case, the solutions can possess both positive and negative net electric charge. When we set $n=-1$, the solution becomes the exact JZ anti-dyon solution and the pure gauge term disappear from the gauge potentials (\ref{eq.25}).

We would also like to comment that the exact JZ anti-dyon solution can be obtained directly from ansatz (\ref{eq.10}) by just setting $R_1$, $R_2$, and $\chi_2$ to zero and $n=-1$.

Finally, we would like to conjecture that a monopole is always located at a point in the Higgs field where the Higgs unit vector $\hat{\Phi}^a$ becomes indeterminate. This point could be a point singularity in the Higgs field or a zero of the Higgs field. We also observed that a point singularity in the Higgs field corresponds to the presence of a Wu-Yang type monopole and a point zero of the Higgs field corresponds to the presence of a 't Hooft-Polyakov monopole of finite energy.

\section*{Acknowlegements}
The author, Rosy Teh, would like to thank Universiti Sains Malaysia and the Academy of Sciences Malaysia for the Scientific Advancement Grant Allocation, SAGA, (Account No.: 304/pfizik/653004/A118). She would also like to thank Prof. Izumi Tsutsui for reading through the manuscript and the Institute of Particle and Nuclear Studies, High Energy Accelerator Research Organization (KEK) Tsukuba, Ibaraki 305-0801, Japan for their hospitality.

\newpage

\end{document}